\documentclass[5p,final]{elsarticle}
\usepackage{amsfonts}
\usepackage{amssymb}
\usepackage{amsmath}  
\usepackage{amsthm}
\usepackage{enumitem} 
\usepackage{geometry}
\usepackage{graphicx}% Include figure files
\usepackage{algorithm}
\usepackage{algcompatible}
\usepackage{algpseudocode}
\usepackage[colorlinks=true, allcolors=blue]{hyperref}
\usepackage{hyperref}
\usepackage[utf8]{inputenc} 
\usepackage[english]{babel} 
\usepackage{url}
\usepackage{wrapfig}
\usepackage{graphicx}
\usepackage{subfiles}
\usepackage{xfrac}
\usepackage{multirow}
\usepackage{cleveref}
\usepackage{balance}
\usepackage{svg}
\usepackage{booktabs}
\usepackage{tikz}
\usetikzlibrary{shapes.geometric, arrows}
\usepackage{tikz-3dplot}
\usepackage[squaren]{SIunits}
\usepackage{caption}
\usepackage{natbib}

\usepackage[normalem]{ulem}
\usepackage{subcaption}
\usepackage{lipsum}
\usepackage{todonotes}
\usepackage{float}
\usepackage{tocloft}
\newcommand{\bc}{\begin{center}}
\newcommand{\ec}{\end{center}}
\newcommand{\be}{\begin{equation}}
\newcommand{\ee}{\end{equation}}
\newcommand{\bea}{\begin{eqnarray}}
\newcommand{\eea}{\end{eqnarray}}
\newcommand{\beq}{\begin{eqnarray*}}
\newcommand{\eeq}{\end{eqnarray*}}
\newcommand{\bv}{\left( \begin{array}{c} }
\newcommand{\ev}{\end{array} \right) }

\usepackage{natbib}
\begin{document}
\title{Non-unique time and market incompleteness}
\author[unsw-mathstats]{Chris Angstmann}\ead{c.angstmann@unsw.edu.au}
\author[uct-sta]{Tim Gebbie} \ead{tim.gebbie@uct.ac.za}
\address[unsw-mathstats]{School of Mathematics and Statistics, University of New South Wales, Sydney, Australia}
\address[uct-sta]{Department of Statisical Sciences, University of Cape Town, Rondebosch 7701, Western Cape, South Africa}
% \email{}

\begin{abstract}
 Financial markets are often modelled as if time were unique and continuous across assets and markets. Financial markets are however asynchronous, order flow is event-driven, and waiting times between events are often random. Many of the most influential formulations of financial market models presuppose a unique global calendar time and advocate for this or that preferred single latent continuous-time price system. Here we critically contrast these assumptions with event-time, renewal, point-process, and order-flow descriptions. We revisit no-arbitrage, no-dynamic-arbitrage, and risk-neutral option pricing in settings where the market is represented as a discrete event system and where the continuum limit of a discrete-time random walk need not be unique. The central suggestion is then that such non-uniqueness points to a more foundational form of market incompleteness than is usually emphasized. This highlights the importance of operational time at the level of decision making but reminds market practitioners that managing risk itself often requires reconciling operational time with a global calendar time. At these longer time scales forms of effective or average completeness may still emerge at lower frequencies and remain useful for portfolio construction and risk management, even if high-frequency hedging and execution expose a clock mismatch between trading, pricing, and longer-horizon allocation.
\end{abstract}

\begin{keyword}
event-time \sep calendar-time \sep discrete time random walks \sep Markov lattice dynamics \\
{\it Subject Areas:} 
%\PACS 89.65.Gh \sep 02.50.Ey
\MSC 82C41 91G60 91G60 91B26
\JEL 91G70
%{\it ACM:} Finance; Electronic Trading; Data Stream Mining; Agent-Based Modeling
% https://cran.r-project.org/web/classifications/JEL.html
\end{keyword}

\maketitle
\tableofcontents

\section{Introduction} \label{sec:intro}

A large part of modern mathematical finance is organized around a single calendar-time price process, usually represented as a semi-martingale adapted to a filtration ${\cal F}_t$ indexed by $t \ge 0$. This representational approach supports stochastic integration, dynamic hedging, and the standard no-arbitrage to martingale-measure logic associated with the fundamental theorem of asset pricing (FTAP). 

In parallel, market microstructure and high-frequency econometrics have developed an event-driven view in which the primitive objects are trades, quotes, order submissions, cancellations, queue dynamics, and the random durations between events. The practical tension between these two views is now familiar, but its conceptual implications are perhaps not as well understood. The issue is whether markets are best understood as a price processes observed irregularly in calendar time, or as a discrete event systems from which prices emerge only after aggregation, synchronization, and time change operations.

The historical lineage for this problem runs from Clark's stochastic business time in the early 1970s \cite{Clark1973}, through the mixture-of-distributions and transaction-time literature of the 1970s and 1980s \cite{Copeland1976,EppsEpps1976,TauchenPitts1983,Harris1987}, to the duration, point-process, and Hawkes-process literature of the 1990s and 2000s \cite{EngleRussell1998,AneGeman2000, Bowsher2007,BacryMastromatteoMuzy2015}, and then to finance CTRW models \cite{MasoliverMonteroWeiss2003,MasoliverMonteroPerelloWeiss2006,ScalasGorenfloMainardi2004,MeerschaertScalas2006} and recent DTRW constructions \cite{AngstmannDonnellyHenryNichols2015}. 

To set the scene we begin by making a strict distinction between ontology (how we chose to represent models) and epistemology (what observations and measurement tells us about what is being modelled). Ontology concerns what is taken to be primitive in a model of financial markets: a calendar-time price process, or a discrete event sequence. This is the choice of modelling framework. Epistemology is then concerned with what can be inferred from the data and how strongly one is entitled to treat observations {\it e.g.} high-frequency records, as samples from a unique latent continuous-time system. This is then about understanding what in fact can be inferred from the data. This separation is useful because many debates in empirical finance conflate the object being modelled with the measurement approach used to estimate it.

\section{The choice of modelling frameworks}
\subsection{The price-first representation}
In the classical price-first ontology or representation, the primitive object is a price process $S=(S_t)_{t\ge 0}$, often taken to be a semi-martingale, It\^o diffusion, or jump-diffusion. This tradition begins with Bachelier and reaches a canonical modern form in \citet{BlackScholes1973} and the martingale approach of Harrison and Kreps \citet{HarrisonKreps1979}, \citet{HarrisonPliska1981}, \citet{DalangMortonWillinger1990}, and \citet{DelbaenSchachermayer1994}. In this setting, calendar time is baked into the filtration and into the admissible class of trading strategies. The attractions of the framework are well known: it supports stochastic integration, self-financing strategies, replication arguments, and state-price deflators or equivalent martingale measures \citep{Bachelier1900,BlackScholes1973,Merton1973,HarrisonKreps1979,HarrisonPliska1981,DalangMortonWillinger1990,DelbaenSchachermayer1994}.

The price-first ontology is powerful, but it is also a modelling commitment. It presupposes that the relevant economic state can be expressed as a process evolving on a single time axis and that the market's observable irregularity can be treated either as noise, as a sampling issue, or as a perturbation around the latent calendar-time evolution. This is often a useful approximation, especially at lower frequencies, but it is not conceptually neutral as it forces a specific ontology.

\subsection{The event-first representation}

In the event-first ontology or representation, the primitive objects are orders, trades, quotes, cancellations, and the matching rules through which market states change. Price is emergent: it is a derived object summarizing the interaction of supply, demand, and liquidity under the market's execution protocol. This view is common in market microstructure and in modern point-process descriptions of limit order books and event arrivals \citep{LilloFarmer2004,TothEtAl2011,FarmerEtAl2013,DonierEtAl2015,BacryMastromatteoMuzy2015}.

The historical roots of the event-first view in finance are older than the recent order-book literature. Clark's \cite{Clark1973} subordination model already suggested that calendar time is not the economically relevant clock for speculative prices. Time subordination changes both scaling and dependence properties of limiting price processes \footnote{Randomising the trading clock changes return distributions: Clark's business-time subordination model gives a finite-variance example. In finance-oriented CTRW models, heavy-tailed waiting times can produce fractional, non-Markovian calendar-time dynamics, so the key modelling choice is the operational (business) time rather than calendar time itself \cite{MainardiRabertoGorenfloScalas2000,Scalas2006,MeerschaertScalas2006}.}. Subsequent work by \citet{EppsEpps1976}, \citet{Copeland1976}, \citet{TauchenPitts1983} and \citet{Harris1987} turned explicitly to transaction volume, transaction counts, and sequential information arrival as mechanisms linking event activity to return distributions and volatility \citep{Clark1973,EppsEpps1976,Copeland1976,TauchenPitts1983,Harris1987}. What changes in later microstructure work is that the event system itself becomes the primary explanatory object rather than merely a latent variance-mixing device (in some imposed calendar time representation).

\section{What is inferred from High-Frequency Data?}

High-frequency financial data are discrete, irregularly spaced, and typically asynchronous across assets and venues. Yet a substantial part of the econometric literature treats such data as non-synchronous observations of a unique latent continuous-time system, developing estimators that aim to recover integrated covariance, lead--lag structure, or other latent calendar-time characteristics despite the fundamental nature of both the noise and asynchrony \citep{HayashiYoshida2005,AitSahaliaFanXiu2010,ChristensenKinnebrockPodolskij2010,HoffmannRosenbaumYoshida2013}.

That project is mathematically sophisticated and empirically useful, but it is not epistemologically clean nor innocent. To infer a latent calendar-time covariance is already to assume that such an object exists and is primary. A different reading of the same data is possible: one may regard measured covariance and correlation as scale-dependent observables whose values depend on synchronization conventions, aggregation schemes, and the choice of sampling clock. On that reading, {\it e.g.} the Epps effect is not merely a nuisance bias to be corrected; it is evidence that dependence itself is clock-relative at short horizons \citep{Epps1979,Chang2020,Chang2021,ChangPienaarGebbie2021,ChangPienaarGebbie2025}.

The distinction matters because the object of inference changes. If one assumes a unique latent instantaneous covariance in calendar time, specialized estimators become recovery devices. If one treats short-horizon dependence as emergent under aggregation over events, trades, or volume, then the target of inference is no longer unique prior to the choice of representation. The empirical literature does not force one reading over the other. The choice is partly inferential and partly ontological. Why this is important is because the direction of the mapping matters; when one is averaging over events to an emergent concept of calendar time, the mapping is unique but explicitly averaged; it is not unique the other way around. 

\section{Stochastic Time in Finance}

\subsection{The business-time hypothesis}

Clark proposed that speculative price changes should be modelled as increments of a process evolving in a stochastic business time rather than in uniform calendar time. In his empirical specification, the random clock was linked to trading activity, and the resulting finite-variance mixtures fit cotton futures returns better than stable-Paretian alternatives \citep{Clark1973}. Clark made two fundamental representational choices. First, the relevant time index for price formation may be stochastic. Second, the empirical fat-tailedness of returns can arise from randomness in the rate of economically relevant activity rather than from infinite-variance innovations.

That basic insight was that observable return distributions reflect mixtures over latent information-arrival or activity states. This  generated the later mixture-of-distributions hypothesis (MDH). Here the crucial point is that the notion of clock relativity is not new; it is already implicit in Clark's rejection of a unique uniform market clock.

\subsection{From volume mixtures to transaction time}

The first major extensions of Clark came quickly. \citet{EppsEpps1976} provided theoretical and empirical support for using transaction volume as a mixing variable for security price changes, showing that the variance of returns could be conditioned on trading activity within a finite-variance mixture framework. \citet{Copeland1976} argued for a sequential information-arrival model that made the pathwise randomness of price adjustments explicit: even if a single piece of information determines the initial-to-final equilibrium price change, the order in which agents receive and act on that information makes both the price path and the volume of trading random. \citet{TauchenPitts1983} deepened the price-variability/volume relation by deriving the joint distribution of daily price changes and volume in speculative markets, while \citet{Harris1987} carried the MDH directly into transaction data, arguing that transaction counts can proxy for latent stochastic variance.

This literature matters because it shows that the finance community has long recognized multiple potential clocks: calendar time, transaction count, and trading volume. What is less often emphasized is that these clocks are not necessarily equivalent. The choice of clock changes the effective distribution of returns, the dependence structure, and the interpretation of information flow. That is what we are concerned with because of it implications for fundamental models of stock markets that arise from order-flow first. 

\subsection{Transaction clocks and uniqueness}

By the late 1990s and early 2000s, the stochastic-clock literature had become more explicit. \citet{AneGeman2000} argued that the cumulative number of trades is a better stochastic clock than volume for recovering approximate normality of returns and emphasized that general stochastic time changes are less restrictive than classical subordinators. Their work is important because it makes clock choice an empirical object rather than a purely theoretical convenience. At the same time, the claims of a unique superior clock were challenged. \citet{MurphyIzzeldin2006} argued that the nonparametric recovery of higher moments of the latent time change in the \citet{AneGeman2000} framework is unstable and that conditioning on transaction counts does not reliably Gaussianize returns in the strong sense often claimed. The lesson is that the clock choice is itself part of the model and may be unidentifiable from data and suggests a representation-level incompleteness.

\section{Duration and Point-Process Approaches}
\subsection{Autoregressive conditional durations}
The duration literature provides a second major genealogy for event-time modelling in finance. \citet{EngleRussell1998} autoregressive conditional duration (ACD) model treats the waiting time between events as a stochastic process with its own dynamics, thereby moving beyond auxiliary proxies for information flow to a direct econometric treatment of irregular event arrivals. 

ACD models allow duration clustering, deterministic intraday seasonality, and duration dependence in a way that is naturally aligned with transaction data. Crucially, they make it possible to model not only price changes but also the arrival intensity of trades and quote revisions. This is important because ACD representations show that irregular timing is not merely an inconvenience to be pre-averaged away; it is itself a stochastic object of interest. The market clock is therefore endogenous to the econometric model.

\subsection{Trade duration and price formation}
\citet{DufourEngle2000} linked event timing directly to price formation, and showed that the price impact of a trade, the speed of price adjustment, and the autocorrelation of signed trades all depend on the duration between consecutive transactions. Shorter inter-trade durations are associated with larger price impact and lower effective liquidity. This is a bridge between event-time econometrics and microstructure theory. 

It implies that the market's informational and liquidity state cannot be represented solely by trade signs and sizes; the temporal spacing of trades is itself informative. If the duration between trades affects price impact, then a representation that suppresses event timing in favour of equally spaced calendar-time returns may misstate the causal structure of price formation.

\subsection{Multivariate point processes and Hawkes-type models}

A more general event-first framework arises in point-process econometrics. \citet{Bowsher2007} developed multivariate intensity-based models for the timing of trades and mid-quote changes, emphasizing that the conditioning information set is updated continuously as market events occur. In parallel, the more recent Hawkes-process literature has shown how self- and cross-excitation can model clustering, mutual feedback between event types, market stability, and even full order-book dynamics \citep{BacryMastromatteoMuzy2015}. 

In these models, prices are no longer a single primitive state variable but one component of a multivariate event system. Once the market is represented as a marked point process or as a mutually exciting event system, the primitive notion of time is the arrival structure itself. Calendar time becomes one external parametrization of event occurrence, not automatically the unique state index.

\section{CTRW, DTRW and Non-Unique Limits}

\subsection{CTRW in finance}
The Continuous-Time Random Walk (CTRW) literature provides the most explicit stochastic-process bridge between event-time models and anomalous calendar-time dynamics. In CTRW models, price changes are represented as jumps separated by random waiting times. \citet{MainardiRabertoGorenfloScalas2000} developed this perspective for finance, deriving waiting-time distributions and showing how non-exponential durations imply non-Markovian effects and, in suitable scaling limits, fractional diffusion equations \citep{MainardiRabertoGorenfloScalas2000,GorenfloMainardiScalasRaberto2001}. However, some care is prudent here because ``fractional'' random walks arise in several non-equivalent senses in the high-frequency finance literature \footnote{In this literature, ``fractional'' may describe the waiting-time law, the limiting evolution equation, or the discrete approximation scheme. In finance-oriented CTRW work it usually refers to heavy-tailed waiting times (and related jump--waiting constructions) whose scaling limits can produce time-fractional master or diffusion equations \cite{ScalasGorenfloMainardi2000,MainardiRabertoGorenfloScalas2000,ScalasGorenfloMainardi2004,MeerschaertScalas2006}. By contrast, \cite{MasoliverMonteroWeiss2003,MasoliverMonteroPerelloWeiss2006} use CTRW primarily to model return distributions and market microstructure, whereas DTRW schemes are useful here because distinct discrete constructions can share the same formal fractional limit while representing different stochastic approximations \cite{AngstmannDonnellyHenryNichols2015,NicholsHenryAngstmann2018}.}

Empirical studies by \citet{RabertoScalasMainardi2002} and theoretical work by \citet{ScalasGorenfloMainardi2004,MasoliverMonteroWeiss2003,MasoliverMonteroPerelloWeiss2006,MeerschaertScalas2006} and others demonstrated that CTRWs seem to offer a natural language for tick-by-tick financial data, where both returns and waiting times matter. However, we are cautioning that there are some implicit choices being made in the time-representations because of the implied latent choices implicit in the global continuum calendar-time representations. 

Three points from the finance CTRW literature are especially relevant here. First, event-time and calendar-time descriptions are linked by a random counting process. Second, the calendar-time process induced by a simple event-time walk need not be diffusive in the classical Brownian sense. Third, dependence between waiting times and jumps matters empirically and can change the limiting behaviour.

\subsection{DTRW and non-uniqueness}

The Discrete-Time Random Walk (DTRW) literature makes a key methodological point that is often underemphasized in finance -- that one can study event systems without needing to pass into a continuum limit. A DTRW is a discrete-time stochastic process whose generalized master equation has, under scaling, the same diffusion or fractional-diffusion limit as the corresponding CTRW. However, the DTRW and CTRW differ in how continuum limits are obtained and interpreted \footnote{Different DTRW constructions may share the same fractional diffusion limit while differing pathwise and in weak approximation \cite{AngstmannDonnellyHenryNichols2015,NicholsHenryAngstmann2018}. For standard uncoupled CTRWs, however, the heavy-tailed Montroll--Weiss scaling leads to a canonical fractional diffusion limit once the tail index and jump scaling are fixed \cite{ScalasGorenfloMainardi2004,GorenfloMainardiScalasRaberto2001}.}.

The advantage is not only numerical. \citet{AngstmannDonnellyHenryNichols2015,AngstmannEtAl2016,NicholsHenryAngstmann2018} show explicitly how the discretization itself is a well-defined stochastic process; so one can study the event system before passing via limits and averaging to any continuum approximation.

This is important because it makes explicit that the map from discrete event dynamics to a continuum equation is not generally unique. Different waiting-time laws, scaling relations, and synchronization schemes can yield different effective continuous-time representations. Brownian diffusion is one possibility, but subordinated Brownian motion, jump processes, and fractional diffusion equations are others. In that sense, the event-to-continuum map is not canonical.

For finance, the implication is not that one must adopt DTRW models literally. Rather, DTRW highlights that a continuous-time representation may be a useful reduced form without being uniquely determined by the underlying discrete event system. This observation is central to the claim of representation-level incompleteness advanced below. In the situation where observations support a discrete event driven foundational reality {\it e.g.} from order-flow models \cite{LilloFarmer2004}, this suggests that the mapping to any continuous representation cannot be unique.  

\section{No-Arbitrage}

\subsection{No-arbitrage}

The economic content of no-arbitrage is older and more general than any one continuous-time representation. In discrete time, the Dalang--Morton--Willinger theorem and its extensions show that no-arbitrage can be formulated directly in terms of predictable strategies and finite stochastic sums, without appeal to stochastic integration in continuous time \citep{DalangMortonWillinger1990,KabanovStricker2006,Bouchard2006}. This is enough to establish the viability principle that zero-cost strategies cannot generate non-negative terminal wealth that is strictly positive with positive probability.

The same observation extends naturally to event time. If the market is indexed by events $n=0,1,2,\ldots$, then a predictable strategy is defined with respect to the event filtration and gains are finite sums over event-time price increments. The economic principle of no-arbitrage is therefore perfectly intelligible before one imposes a global calendar-time based semi-martingale structure.

\subsection{No-dynamic-arbitrage}

A similar remark applies to no-dynamic-arbitrage in market impact models. The standard continuous-time no-price-manipulation condition, as developed by \citet{HubermanStanzl2004} and by \citet{Gatheral2010}, prohibits profitable round trips under a specified impact kernel and execution model. But the principle itself does not require calendar time as primitive or foundational time representation. One can formulate an event-time version in which predictable trade increments over the event sequence must not generate negative expected total execution cost for a round trip. Again, the viability principle survives the change of ontology (or representation).

The strongest criticism, then, should not be directed at no-arbitrage itself but at any tendency to identify it too closely with one mathematical representation of time.

\subsection{Risk-neutral pricing}

Risk-neutral pricing in the classical sense links no-arbitrage to the existence of an equivalent martingale measure, uniquely in complete models and non-uniquely in incomplete ones \citep{BlackScholes1973,Merton1973,HarrisonKreps1979,HarrisonPliska1981,DelbaenSchachermayer1994}. In a discrete event-time setting, analogous martingale pricing results remain available under suitable assumptions \citep{DalangMortonWillinger1990,Bouchard2006}. What changes under clock relativity is not the logic of arbitrage-free pricing per se but the level at which representation is fixed.

If different event-to-continuum maps support different effective continuous-time models, then a pricing measure is attached to a chosen representation rather than to the market in some absolute sense. Even before one reaches the usual incomplete-market multiplicity of equivalent martingale measures, there may already be inescapable non-uniqueness in the underlying state description itself. This motivates the claim that incompleteness can be more foundational than the standard spanning-based perspectives suggest.

\section{Representation-Level Incompleteness}

The Epps effect is a natural diagnostic for the issues raised above. Measured cross-asset correlations decline as the sampling interval is shortened, and the standard explanations invoke asynchronous trading, lead--lag relations, and microstructure frictions \citep{Epps1979,HayashiYoshida2005,AitSahaliaFanXiu2010}. One interpretation is that a unique latent instantaneous correlation exists in calendar time and that specialized estimators can recover it despite imperfect sampling. Another interpretation is that dependence itself is scale dependent and clock dependent, so that no single instantaneous correlation parameter deserves ontological priority at all scales \citep{Chang2020,Chang2021,ChangPienaarGebbie2021,ChangPienaarGebbie2025}.

The second interpretation fits more naturally with the event-time, duration, and DTRW approach (and by implication aspects of the CTRW approach reviewed above -- but it is important to be aware of potential problems with no-arbitrage\footnote{In finance, the objection to CTRW is not that it is generically non-semimartingale; rather, CTRW’s fractional continuous-time limits (which introduce random calendar-time waiting and then studies diffusion/fractional limits) can sit awkwardly with standard continuous-trading no-arbitrage machinery, whereas DTRW/event-time models preserve the native discreteness of trading and fit more naturally with discrete-time arbitrage frameworks \cite{HarrisonKreps1979, DalangMortonWillinger1990, Bouchard2006} within an event-first representation.}). If trade arrivals are irregular and if event sequences differ across assets, then synchronization is not merely a statistical nuisance; it is part of the mechanism through which dependence becomes observable. A correlation measured in calendar time, trade time, or volume time need not be a noisy estimate of the same primitive object.

This motivates the notion of \emph{representation-level incompleteness}. In the usual complete/incomplete market dichotomy, incompleteness means that not every contingent claim can be replicated within a fixed state description. Here the suggestion is stronger: the market may not determine a unique state description or a unique clock under which hedging, covariance measurement, and pricing are jointly canonical. If so, incompleteness appears one level earlier, at the level of representation.

\section{Effective Completeness and Emergence}

None of the preceding implies that semimartingale finance is invalid or useless. On the contrary, coarse-grained calendar-time models often work well for lower-frequency portfolio construction, factor allocation, and benchmark-relative risk management. The practical success of such models suggests that some degree of \emph{effective completeness} or at least effective stability can emerge under aggregation even when the underlying event system is richer and more irregular \citep{HarrisonPliska1981,DelbaenSchachermayer1994}.

The important point is that such completeness may be emergent rather than universal. Trading desks, option hedgers, and funding desks operate close to execution time and liquidity conditions, whereas asset-management and mutual-fund risk systems operate on aggregated horizons. If there is no unique clock bridge linking these levels, then the disconnect may manifest not as literal arbitrage but as slippage, basis risk, model dependence, and state-dependent funding or liquidity costs \citep{CetinJarrowProtterWarachka2006,BrunnermeierPedersen2009}. The possibility that forward and futures structures can reflect these state-dependent distortions also fits naturally into this picture \citep{JarrowProtterShimbo2010}.

\section{Conclusion}

Financial market theory should distinguish more sharply between: (i) viability principles such as no-arbitrage and no-dynamic-arbitrage; (ii) the clock and filtration under which strategies are defined; and (iii) the scale at which effective completeness is claimed to hold. Once the market is treated as a discrete event system with multiple admissible clocks, the move from event time to continuum time is not generally unique. This does not eliminate arbitrage-free pricing. It does, however, make such pricing explicitly model-relative and clock-relative unless further structure is imposed.

If that is right, then market incompleteness may be more foundational than is commonly assumed. A complete market may still emerge as a useful low-frequency approximation, but not necessarily as a universal structure that connects seamlessly across clocks, scales, and market functions.

\bibliography{time}

%% BIBTEX
\appendix
\end{document}